\documentclass[10pt, conference, letterpaper, twocolumn, twoside]{IEEEtran}

\usepackage{cite}
\usepackage{graphicx}
\usepackage{subfigure}
\usepackage{amsmath}
\usepackage{epsfig}
\usepackage{times}

\newtheorem{pavikl}{\em Lemma}
\newtheorem{pavikt}{\em Theorem}

\begin{document}

\title{Energy Conscious Interactive Communication for Sensor Networks}
\author{\authorblockN{{\Large Samar Agnihotri and Pavan Nuggehalli}}\\
\authorblockA{Centre for Electronics Design and Technology, Indian Institute of Science, Bangalore - 560012, India\\
Email: \{samar, pavan\}@cedt.iisc.ernet.in}%
}

\maketitle

\begin{abstract}
In this work, we are concerned with maximizing the lifetime of a cluster of sensors engaged in single-hop communication with a base-station. In a data-gathering network, the spatio-temporal correlation in sensor data induces data-redundancy. Also, the interaction between two communicating parties is well-known to reduce the communication complexity. This paper proposes a formalism that exploits these two opportunities to reduce the number of bits transmitted by a sensor node in a cluster, hence enhancing its lifetime. We argue that our approach has several inherent advantages in scenarios where the sensor nodes are acutely energy and computing-power constrained, but the base-station is not so. This provides us an opportunity to develop communication protocols, where most of the computing and communication is done by the base-station.

The proposed framework casts the sensor nodes and base-station communication problem as the problem of multiple informants with correlated information communicating with a recipient and attempts to extend extant work on interactive communication between an informant-recipient pair to such scenarios. Our work makes four major contributions. Firstly, we explicitly show that in such scenarios interaction can help in reducing the communication complexity. Secondly, we show that the order in which the informants communicate with the recipient may determine the communication complexity. Thirdly, we provide the framework to compute the $m$-message communication complexity in such scenarios. Lastly, we prove that in a typical sensor network scenario, the proposed formalism significantly reduces the communication and computational complexities.
\end{abstract}

\section{Introduction}
\label{Intro}
A typical wireless sensor network consists of sensor nodes with limited energy reserves. Many sensor network applications expect the sensor nodes to be active for months and may be years. However, in most of the situations, once the sensor nodes run out of their energy reserves, then replacing their batteries is not possible either due to the inaccessibility of sensor nodes or because such an endeavor may not be economically viable. So, there is a great demand and scope of the strategies which attempt to reduce the energy consumption, hence increase the lifetime of the sensor nodes.

Sensor nodes spend energy in receiving and transmitting data, sensing/actuating, and computation. In this paper, we concern ourselves with reducing the energy cost of transmission. The energy cost of receiving data can be easily incorporated in the model that we propose. The energy spent in sensing/actuating represents a fixed cost that can be ignored. We assume computation costs are negligible compared to radio communication costs. Though this is a debatable assumption in dense networks, but incorporating computation costs is not straightforward and is left for future work.

The transmission energy depends on three factors: the number of bits to be transmitted, the path-loss factor between the sensor nodes and the base-station, and the time available to transmit the given number of bits. The path-loss factor describes the wireless channel between a sensor node and the base-station and captures various channel effects, such as distance induced attenuation, shadowing, and multipath fading. For simplicity, we assume the path-loss factor to be constant. This is reasonable for static networks and also the scenarios where the path-loss factor varies slowly and can be accurately tracked. The idea of varying the transmission time to reduce the energy consumption was proposed in \cite{Prabhakar} and explored in \cite{wowmom05} in the context of sensor networks, where its $\cal{NP}$-hardness is discussed. In this paper, we attempt to reduce the transmission energy by reducing the number of bits transmitted by the sensor nodes to the base-station.

In a data gathering sensor network, the spatio-temporal correlations in sensor data induce data-redundancy. In \cite{Slepian}, Slepian and Wolf show that it is possible to compress a set of correlated sources down to their joint entropy, without explicit communication between the sources. This surprising existential result shows that it is enough for the sources to know the probability distribution of data generated\footnote{Actually, the knowledge of conditional entropies suffices.}. Recent advances \cite{Pradhan} in distributed source coding allow us to take advantage of data correlation to reduce the number of bits that need to be transmitted, with concomitant savings in energy. However, finding the optimal rate allocation lying in Slepian-Wolf achievable rate region defined by $2^N-1$ constraints for $N$ nodes and designing efficient distributed source codes is a challenging problem. We simplify this problem by allowing the interaction between the base-station and the sensor nodes, and introducing the notion of \textit{instantaneous decoding} \cite{wowmom05}. This reduces the rate allocation problem to the problem of finding the optimal scheduling order, albeit at some loss of optimality.

Two results in the theory of interactive communication provided further motivation for our work. First result states that even if the feedback does not help in increasing the capacity of a communication channel, it does help in reducing the complexity of communication \cite{Cover_book, Cover_feedback}. Second result states that for the worst-case interactive communication, for almost all \textit{sparse pairs} the recipient, who has nothing to say, must transmit almost all the bits in an optimal interactive communication protocol and informant transmits \textit{almost} nothing \cite{090orlitsky}.

Based on these results, this paper proposes a formalism that attempts to reduce the number of bits sent by a sensor node. The proposed formalism casts the problem of many sensor nodes communicating with a base-station as a problem where multiple informants with correlated information communicate with single recipient. Here we identify the base-station as the recipient of the information and the sensor nodes as the sources of information. However, \cite{090orlitsky} and subsequent papers \cite{091orlitsky, 092orlitsky, 094zhang, 097ahlswede} consider only the interactive communication between a single informant-recipient pair, while in the sensor networks, we have as many informant-recipient pairs, as the number of sensor nodes. If the sensor data is assumed to be uncorrelated, then the results of \cite{090orlitsky} can be trivially extended to the present scenario. However, in a data-gathering sensor network, the sensor data is supposed to be correlated. So, extending the results in \cite{090orlitsky, 092orlitsky, 097ahlswede} to the scenarios where multiple informants with correlated information communicate with single recipient and then applying those results to the sensor networks, is not straightforward.

Formally, the problem is the following. There are $N$ sources of correlated information and there is one recipient of information $\cal P$, which needs to collect the information from these $N$ sources. Assume that $\cal P$ can interact with any of those $N$ sources, but among themselves these $N$ sources cannot interact directly. At the end of communication, each of these $N$ sources need not know what other $N-1$ sources know. We are looking for the most efficient communication schemes (ones which minimize the communication complexity) for this problem. In the context of sensor networks, if the base-station knows the joint distribution of the sensor data or the correlations therein, then it can play some distributed version of the `game of twenty questions' with $N$ sensors to retrieve their information. If there is a single sensor node or if the sensor data are uncorrelated, then the base-station can `play' with individual sensor nodes and any node $i$ needs to send at least $H_i(X)$ bits, where $H_i(X)$ is the entropy of the information source at node $i$. So, assuming the sensor data at all the nodes are \textit{identically distributed}, finally the base-station retrieves $N H(X)$ bits from the $N$ sensors. However, in a more realistic situation, the sensors have correlated data and the base-station needs to retrieve only $H(X_1, ... X_N) \le N H(X)$ bits. Note that we can talk in terms of entropies only when we are concerned with average number of bits communicated. However, if we are interested in the worst-case number of bits, then it depends solely on the cardinality of the \textit{support set} of the data of individual nodes than on the corresponding probabilities.

In this work, we make two simplifying assumptions: the sensor nodes communicate with the base-station in a single-hop and only the total number of bits exchanged between the sensor nodes and the base-station are considered and this is what we refer to as `communication complexity'. This paper shows that the communication complexity depends on the model of the spatial-correlation of the sensor data as well as on the order in which the sensor nodes communicate with the base-station. Under the assumption of an omniscient base-station, the base-station can compute the optimal number of bits, which any node in any schedule needs to send to it for any spatial correlation model of the sensor data. However, a sensor node, even if it knows in how many bits it needs to send its information to the base-station, may not actually be able to compress its information to that many number of bits, without explicit knowledge of how its data is correlated with the data of the other nodes. In general, this is neither possible given the limited knowledge of the network that a sensor node is supposed to have, nor desirable given the limited energy and computational capabilities of the sensor nodes. An omnipotent base-station can take up the most of the burden of computation and communication, allowing the the sensor nodes to perform minimal computation and transmit minimal number of bits, hence conserving their precious energy reserves and possibly increasing their operational lifetime.

In the section \ref{Example}, we give an example of the situation where multiple informants with correlated information communicate with single recipient. We use this example to illustrate the complexity of communication in such scenarios and various other fundamental issues and our results. Section \ref{worst_case_theory} develops the formalism to compute the worst-case communication complexity in the scenarios where multiple correlated informants communicate with a single recipient. Section \ref{sensorNet_appl} illustrates some of the ideas proposed and developed in this paper in the context of a sensor network with one particular model of spatial correlation of sensor data. Finally, the section \ref{conclusions} lists the contributions of our work and concludes the paper.

\section{An Example}
\label{Example}
There are $N$ groups $\{S_i, 1 \le i \le N \}$, and each group has $t$ teams. In every match, two teams from two different groups play against each other. The result of each match is announced over radio. Matches always result in a clear winner. The format of the radio announcements is: ``Today the teams from groups $I$ and $J$ played against each other. The match between teams $i$ and $j$ was won by team $k$.'' Three persons $P_{\cal{X}}$, $P_{\cal{Y}}$, and $P_{\cal{Z}}$ are involved. $P_{\cal{X}}$ listens to the first part of the announcement ``Today the teams from groups $I$ and $J$ played against each other.'' and then the radio is snatched by person $P_{\cal{Y}}$, who listens to the portion ``The match between teams $i$ and $j$ was won by'' before the radio is snatched from him by $P_{\cal{Z}}$, who listens to the portion ``team $k$''. Now all three persons agree that $P_{\cal{X}}$ must know which two teams played and who the winner was ($P_{\cal{Y}}$ and $P_{\cal{Z}}$ need not learn which two teams from which two groups actually played and which one finally won). They want to find the most efficient way to do this.

$P_{\cal{X}}$ wants to know which two teams from the groups $I$ and $J$ played and which one actually won. Suppose only $P_{\cal{Y}}$ communicates with $P_{\cal{X}}$, then $P_{\cal{X}}$ will only know which two teams from the groups $I$ and $J$ actually played the match, but not the winner. On the other hand, if only $P_{\cal{Z}}$ communicates with $P_{\cal{X}}$, then $P_{\cal{X}}$ knows the teams from which two groups played and who was the winner, but he may not know who the other team was. In this problem, it is essential that the identity of the group to which a team belongs to is included in the name of the team, that is, the names of the teams have to be globally unique. Suppose on the contrary, that the names of the teams are only unique within a group, then two or more groups might have the the teams with the same name. So, in the event of a match between two teams from two different groups with the same name playing against each other, $P_{\cal{X}}$ will not be able to make the winner out of the information sent to him by $P_{\cal{Z}}$, even if $P_{\cal{Y}}$ has already informed $P_{\cal{X}}$ of which two teams have played. So, we are demanding that once $P_{\cal{Y}}$ and $P_{\cal{Z}}$ have communicated all their information to $P_{\cal{X}}$, then $P_{\cal{X}}$ must be able to unambiguously infer which two teams had played as well as who the winner was. The previous argument proves that this demand is satisfied if and only if the team names are globally unique.

It should be noted that the total number of bits exchanged to complete the entire process of communication depends very much on the format of the announcement. For example, in the original problem in \cite{090orlitsky}, if the format of the announcement is: ``the match between teams $i$ and $j$ is won by first/second team'' with the protocol that the `first' (`second') corresponds to the first (second) team mentioned in the announcement. In such a situation, only one bit needs to be sent to $P_{\cal{Y}}$, with or without interaction between $P_{\cal{X}}$ and $P_{\cal{Y}}$.

\subsection{Without interaction}
$P_{\cal{Y}}$ sends $\lceil \log N \rceil + \lceil \log t \rceil$ bits to identify one of the groups and the team from it. After this message, it does not need to identify the second group as it is obvious to $P_{\cal{X}}$. So, $P_{\cal{Y}}$ needs to send another $\lceil \log t \rceil$ bits to enable $P_{\cal{X}}$ know which is the team from the other group (message 1, 2). $P_{\cal{Z}}$ sends $\lceil \log N + \log t \rceil$ bits to $P_{\cal{X}}$ to help him know who the winner was (message 3).
\begin{eqnarray}
\label{no_interaction}
P_{\cal{X}} \mbox{ sends} &:&  \mbox{nothing}\\
P_{\cal{Y}} \mbox{ sends} &:& \lceil \log N \rceil + \lceil \log t \rceil + \lceil \log t \rceil \nonumber \\
                  &=& \lceil \log N \rceil + 2 \lceil \log t \rceil \nonumber \\
P_{\cal{Z}} \mbox{ sends} &:& \lceil \log N + \log t \rceil \nonumber
\end{eqnarray}

\subsection{With interaction}
There are two scenarios here. One where $P_{\cal{Y}}$ communicates its information to $P_{\cal{X}}$ before $P_{\cal{Z}}$ communicates with $P_{\cal{X}}$. The other scenario is the one where $P_{\cal{Z}}$ sends his information to $P_{\cal{X}}$ before $P_{\cal{Y}}$ sends. We have to compute the number of bits exchanged for both the scenarios.

\textbf{When $P_{\cal{Y}}$ communicates with $P_{\cal{X}}$ before $P_{\cal{Z}}$:} $P_{\cal{X}}$ knows the two groups from which two teams played against each other. $P_{\cal{X}}$ encodes in $\lceil \log N \rceil$ bits the names of the groups. So, $P_{\cal{X}}$ sends in $\lceil \log \log N \rceil$ bits the first bit location at which the encodings of the two groups differ to $P_{\cal{Y}}$ (message 1). $P_{\cal{Y}}$ on its turn, sends the value of the first bit at which the encodings of two groups differ along with the $\lceil \log t \rceil$ bits to identify the team within one of the groups. To help identify the team from the other group it just needs to send $\lceil \log t \rceil$ bits to $P_{\cal{X}}$ as the identity of the group to which this team belongs to is already known to $P_{\cal{X}}$ by now (message 2, 3, 4). At the end of this step, $P_{\cal{X}}$ knows which two teams had played. So, now in $\lceil \log (\log N + \log t) \rceil$ bits it sends the identity of the first bit location at which the encodings of the two teams differ to $P_{\cal{Z}}$ (message 5) and $P_{\cal{Z}}$ responds by sending the value of that bit (message 6). With this $P_{\cal{X}}$ can determine who the winner was. So under this scenario, the number of bits exchanged are:
\begin{eqnarray}
\label{y_sends_before_z}
P_{\cal{X}} \mbox{ sends} &:& \lceil \log \log N \rceil + \lceil \log (\log N + \log t) \rceil \\
P_{\cal{Y}} \mbox{ sends} &:& 1 + \lceil \log t \rceil + \lceil \log t \rceil = 1 + 2 \lceil \log t \rceil \nonumber \\
P_{\cal{Z}} \mbox{ sends} &:& 1 \nonumber
\end{eqnarray}

\textbf{When $P_{\cal{Z}}$ communicates with $P_{\cal{X}}$ before $P_{\cal{Y}}$:} $P_{\cal{X}}$ sends in $\lceil \log \log N \rceil$ bits the location of the first bit at which the encodings of the two groups differ to $P_{\cal{Z}}$ (message 1). $P_{\cal{Z}}$ in its turn, sends the value of the bit at that location along with the encoding of the winning team in $\lceil \log t \rceil$ number of bits (message 2, 3). At the end of this step, $P_{\cal{X}}$ knows which was the winning team and to which group it belonged to too. So, all that $P_{\cal{X}}$ does not know now is that which team from the other group also played in the match. It sends to $P_{\cal{Y}}$ the location and value of the first bit at which the encodings of two two teams differ in $1 + \lceil \log \log N \rceil$ bits (message 4, 5). $P_{\cal{Y}}$ then responds by sending $\lceil \log t \rceil$ number of bits to identify the team from the given group (message 6). So under this scenario, the number of bits exchanged are:
\begin{eqnarray}
\label{z_sends_before_y}
P_{\cal{X}} \mbox{ sends} &:& \lceil \log \log N \rceil + 1 + \lceil \log \log N \rceil \\
                  &=& 1 + 2 \lceil \log \log N \rceil \nonumber \\
P_{\cal{Y}} \mbox{ sends} &:& \lceil \log t \rceil \nonumber \\
P_{\cal{Z}} \mbox{ sends} &:& 1 + \lceil \log t \rceil \nonumber
\end{eqnarray}

Note \eqref{no_interaction}, \eqref{y_sends_before_z}, and \eqref{z_sends_before_y} show that if we consider, total number of bits exchanged in the entire communication or the total number of bits transmitted by persons $P_{\cal{Y}}$ and $P_{\cal{Z}}$ together, interaction helps in reducing the number of bits compared to when no interaction is allowed. Further, when $P_{\cal{Z}}$ communicates with $P_{\cal{X}}$ before $P_{\cal{Y}}$, the number of exchanged bits are less than when $P_{\cal{Y}}$ communicates with $P_{\cal{X}}$ before $P_{\cal{Z}}$.

If we adopt the convention that all the messages sent by a source, until some other source sends the messages, form one message, then in all the above situations at most four messages are exchanged. So, in a communication protocol following this convention, the source concatenates all the messages that it sends before some other source begins to send the messages, and receiver knows how to parse the concatenated message into its individual messages.

As \cite{090orlitsky, 091orlitsky, 097ahlswede} and the example above prove, the interaction reduces the number of bits exchanged between the informants and the recipient. However, from the above example, it also becomes clear that even for the given number of messages exchanged between the informants and the recipient, in general, the number of bits exchanged depends on the order in which the informants communicate with the recipient. So in the above example, with four messages allowed, the number of bits exchanged between $P_{\cal{X}}, P_{\cal{Y}}$, and $P_{\cal{Z}}$ depend on whether $P_{\cal{Y}}$ communicates with $P_{\cal{X}}$ before $P_{\cal{Z}}$ or after. We conjecture that this is so due to ``somewhat'' asymmetric nature of the distribution of the information at the nodes $P_{\cal{Y}}$ and $P_{\cal{Z}}$ of this particular example and loosely, we can say that the messages from $P_{\cal{Z}}$ contain more information than those from $P_{\cal{Y}}$.

\section{Worst case interactive communication: multiple informants scenario}
\label{worst_case_theory}
In this section we attempt to develop a theory of worst-case interactive communication. To keep the discussion simple and clear, we consider a general scenario involving two informants ($P_{\cal{Y}}$ and $P_{\cal{Z}}$) and one recipient ($P_{\cal{X}}$). However, the same formalism can be extended for the scenarios involving more than two informants. For the sake of completeness and to facilitate the discussion that follows, let us reintroduce the notion of \textit{ambiguity set} and some related concepts, defined originally in \cite{090orlitsky}.

Let $(X, Y)$ be a random pair, with \textit{support set} $S_{X,Y}$. The \textit{support set} of $X$ is the set
\begin{equation*}
S_X \stackrel{\mbox{def}}{=} \{x: \mbox{ for some } y, (x, y) \in S_{X, Y}\}
\end{equation*}
of possible $X$ values. $S_Y$, the \textit{support set} of $Y$, is similarly defined. $P_{\cal{Y}}$'s \textit{ambiguity set} when his random variable $Y$ takes the value $y \in S_Y$ is
\begin{equation}
\label{eqn:ambiguity_set}
S_{X|Y}(y) \stackrel{\mbox{def}}{=} \{x: (x, y) \in S_{X, Y} \},
\end{equation}
the set of possible $X$ values when $Y = y$. His \textit{ambiguity} in that case is
\begin{equation*}
\mu_{X|Y}(y) \stackrel{\mbox{def}}{=} |S_{X|Y}(y)|,
\end{equation*}
the number of possible $X$ values when $Y=y$. The \textit{maximum ambiguity} of $(X, Y)$ is
\begin{equation}
\label{eqn:max_ambiguity}
\hat{\mu}_{X|Y} \stackrel{\mbox{def}}{=} \sup \{\mu_{X|Y}(y): y \in S_Y\},
\end{equation}
the maximum number of $X$ values possible with any given $Y$ value.

Assume that a total of $2m$ messages are allowed, with $m$ messages per informant allowed to be exchanged between the informant and the recipient. There are two possible schedules in which the informants can communicate with the recipient: either $P_{\cal{Y}}$ communicates first with $P_{\cal{X}}$ or $P_{\cal{Z}}$ communicates first. In the spirit of \cite{090orlitsky}, let us introduce some more definitions:

$\hat{C}_{2m}^{YZ}(Y,Z/X)$: $2m$-message worst-case complexity of transmitting $Y, Z$ to a person who already knows $X$, when $Y$ communicates first.

$\hat{C}_{2m}^{ZY}(Y,Z/X)$: $2m$-message worst-case complexity of transmitting $Y, Z$ to a person who already knows $X$, when $Z$ communicates first.

Using these definitions, we have:
\begin{eqnarray}
\hat{C}_{2m}^{YZ}(Y,Z/X) = \hat{C}_m(Y/X) + \hat{C}_m(Z/X,Y) \\
\hat{C}_{2m}^{ZY}(Y,Z/X) = \hat{C}_m(Z/X) + \hat{C}_m(Y/X,Z)
\end{eqnarray}

\begin{pavikt}
In general, $\hat{C}_{2m}^{YZ}(Y,Z/X) \neq \hat{C}_{2m}^{ZY}(Y,Z/X)$.
\end{pavikt}
\begin{proof}
We omit the detailed proof for the sake of brevity. However, the league example of the previous section provides an example supporting the statement of the theorem.
\end{proof}

\textit{Corollary:} The unbounded interaction complexities satisfy $\hat{C}_{\infty}^{YZ}(Y,Z/X) \neq \hat{C}_{\infty}^{ZY}(Y,Z/X)$.

It is easy to prove the following trivial, but quite useful lower bounds on the unbounded interaction complexities.
\begin{pavikl}
For all (X, Y, Z) tuples,
\begin{eqnarray*}
\hat{C}_{\infty}(Y/X) & \ge & \lceil \log \hat{\mu}_{Y|X} \rceil, \\
\hat{C}_{\infty}(Z/X,Y) & \ge & \lceil \log \hat{\mu}_{Z|X,Y} \rceil, \\
\hat{C}_{\infty}^{YZ}(Y,Z/X) & \ge & \lceil \log \hat{\mu}_{Y,Z|X} \rceil.
\end{eqnarray*}
Similar bounds exists for $\hat{C}_{\infty}(Z/X)$, $\hat{C}_{\infty}(Y/X,Z)$, and $\hat{C}_{\infty}^{ZY}(Y,Z/X)$.
\end{pavikl}

Since empty messages are allowed, it is obvious that $\hat{C}_m(Y/X)$ is a decreasing function of $m$. This holds true for other complexities, such as $\hat{C}_m(Z/X)$ etc too. This fact together with previous lemma implies that
\begin{eqnarray*}
\hat{C}_m(Y/X) & \ge & \lceil \log \hat{\mu}_{Y|X} \rceil, \\
\hat{C}_m(Z/X,Y) & \ge & \lceil \log \hat{\mu}_{Z|X,Y} \rceil, \\
\hat{C}_{2m}^{YZ}(Y,Z/X) & \ge & \lceil \log \hat{\mu}_{Y,Z|X} \rceil.
\end{eqnarray*}
With similar bounds for $\hat{C}_m(Z/X)$, $\hat{C}_m(Y/X,Z)$, and $\hat{C}_{2m}^{ZY}(Y,Z/X)$.

We can use above results to find the communication complexity of the version of the league problem discussed in the previous section and arrive at the results of \eqref{y_sends_before_z} and \eqref{z_sends_before_y}. If we identify the sets $X, Y$, and $Z$ appropriately, we can directly use the results from \cite{090orlitsky}. For example, the relevant support sets $S_{Y, X}$, $S_{Z;X,Y}$, $S_{Z,X}$, and $S_{Y;X,Z}$, with $1 \le i \ne j \le N$ are:
\begin{eqnarray*}
S_{Y, X} & = & \{((k, l), (S_i, S_j)), k \in S_i, l \in S_j \} \\
S_{Z;X,Y} & = & \{ (k, (m, n)), m \in S_i, n \in S_j, k \in \{m, n\} \} \\
S_{Z,X} & = & \{(k, (S_i, S_j)), k \in S_i \cup S_j \} \\
S_{Y;X,Z} & = & \{ (k, S_i), S_i \mbox{ is not the winner's group}, k \in S_i \}
\end{eqnarray*}

\section{Application to Sensor Network}
\label{sensorNet_appl}
In this section, we apply the formalism developed in the previous sections to illustrate the computation of worst-case and average-case interactive communication complexities. We assume following spatial correlation model for the sensor data.

Let $X_i$ be the random variable representing the sampled sensor reading at node $i\in \{1, \ldots, N\}$ and $B(X_i)$ denote the number of bits that the node $i$ has to send to the base-station. Let us assume that each node $i$ has at most $n$ number of bits to send to the base-station, so $B(X_i) = n$. However, due to the spatial correlation among sensor readings, each sensor may send less than $n$ number of bits. Let us define a data-correlation model as follows.

Let $d_{ij}$ denote the distance between nodes $i$ and $j$. Let us define $B(X_i/X_j)$, the number of bits that the node $i$ has to send when the node $j$ has already sent its bits to the base-station, as follows:
\begin{equation}
\label{eqn:cor}
B(X_i/X_j) = \left\{
                    \begin{array}{ll}
                     \lceil d_{ij} \rceil \mbox{ if } d_{ij} \le n\\
                     n \mbox{ if } d_{ij} > n
                    \end{array}
             \right.
\end{equation}

\begin{figure}[t]
\begin{center}
\includegraphics[angle=-90, width=8.85cm]{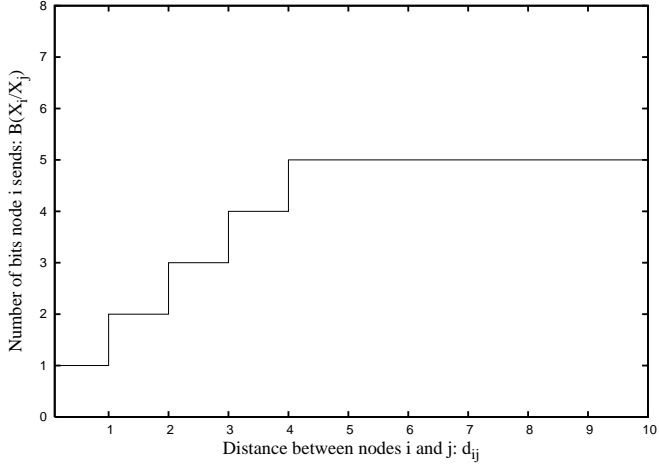}
\end{center}
\caption{Data Correlation Model for $n=5$: plot of $B(X_i/X_j)$ versus $d_{ij}$}
\label{fig1}
\end{figure}

Figure \ref{fig1} illustrates this for $n=5$. It should be noted that when $\lceil d_{ij} \rceil < n$, the data of the nodes $i$ and $j$ differ in at most $B(X_i/X_j)$ least significant bits. So, the node $i$ has to send, at most, $\lceil d_{ij} \rceil$ least significant bits of its $n$ bit data.

From the definition above in \eqref{eqn:cor}, follows the symmetry of the conditional number of bits:
\begin{equation}
\label{eqn:cor_sym}
B(X_i/X_j) = B(X_j/X_i)
\end{equation}

However, the definition of the correlation model is not complete yet and we must give the expression for the number of bits transmitted by a node conditioned on more than one node already having transmitted their bits to the base-station. There are several ways in which this quantity can be defined, we choose the following definition:
\begin{equation}
\label{eqn:cor_cond}
B(X_i/X_1, \ldots, X_{i-1}) = \min_{1 \le j < i} B(X_i/X_j)
\end{equation}

So, with this definition, the number of bits transmitted by any node in a schedule depends only on the node nearest to it among all the nodes already polled in the schedule.

\subsection{Worst-case communication}
For a given schedule $\pi$, first node $\pi(1)$ transmits its $n$ bits to the base-station. The omniscient base-station based on its knowledge of the correlation model in \eqref{eqn:cor} as well as all the internode distances, knows that its \textit{maximum ambiguity} about second node's data is $2^{B(X_{\pi(2)}/X_{\pi(1)})}$, as $2^{B(X_{\pi(2)}/X_{\pi(1)})}$ bit-patterns are possible for the $B(X_{\pi(2)}/X_{\pi(1)})$ least-significant bits of $\pi(2)$'s data. So, even if we allow unbounded interaction between the base-station and the node $\pi(2)$, it follows from the results of the previous section that at least $B(X_{\pi(2)}/X_{\pi(1)})$ bits are exchanged, even by the optimal communication protocol. However, here we are more concerned with demonstrating the reduction in the communication complexity due to the interaction between the sensor nodes and the base-station rather than with proposing the schemes which achieve the optimal lower bounds of the communication complexity. So, here we propose an \textit{almost optimal} protocol for the communication between the base-station and a sensor node. Base-station informs the node $\pi(2)$ in $\lceil \log B(X_{\pi(2)}/X_{\pi(1)}) \rceil$ bits (if $B(X_{\pi(2)}/X_{\pi(1)}) = 1$, then the base-station sends one bit to the corresponding node) to transmit $B(X_{\pi(2)}/X_{\pi(1)})$ least significant bits of its information and the sensor node $\pi(2)$ responds by sending corresponding bits. Note that if no interaction is allowed between the base-station and the sensor node, then the sensor node has to send $n$ bits to the base-station. Continuing this process, the base-station queries all the $N$ nodes. So, the worst-case communication complexity $\hat{C}_{\pi}$ of schedule $\pi$ is the sum of the total number of bits sent by the base-station and the number of bits sent by the sensor-nodes, in the worst-case. It is given by:
\begin{eqnarray}
\label{eqn:worst_case}
\hat{C}_{\pi} & = & \lceil \log B(X_{\pi(2)}/X_{\pi(1)}) \rceil + \ldots \nonumber \\
          & & + \lceil \log B(X_{\pi(N)}/X_{\pi(1)}, \ldots, X_{\pi(N-1)}) \rceil \nonumber \\
          & & + B(X_{\pi(2)}/X_{\pi(1)}) + \ldots \nonumber \\
          & & + B(X_{\pi(N)}/X_{\pi(1)}, \ldots, X_{\pi(N-1)})
\end{eqnarray}

Let $\Pi$ be the set of all possible schedules. Then optimally minimum value of $\hat{C}_{\pi}$ is achieved for that schedule $\pi \in \Pi$ that solves the following optimization problem:
\begin{equation}
\label{optprob1}
\hat{C}_{min} = \min_{\pi \in \Pi} \hat{C}_{\pi}
\end{equation}

Given the definition of the correlation model in \eqref{eqn:cor} and \eqref{eqn:cor_cond}, it is easy to see that the optimum schedule is generated by a greedy scheme that chooses the next node in the schedule (from the set of the nodes not already scheduled) to be that node that is nearest to the set of already scheduled nodes.

\subsection{Average-case communication}
As noted above in the discussion of the correlation model of \eqref{eqn:cor}, the data of two correlated nodes $i$ and $j$ can differ in at most $B(X_i/X_j)$ number of least significant bits. However, it is not necessary that it \textit{actually} differs in those many bits. So, there can be the situations where even if the base-station has estimated that the node $i$ has to send $B(X_i/X_j)$ number of its least-significant bits and communicated this to the node $i$, the node $i$'s data differs from the the data of node $j$ in less than $B(X_i/X_j)$ least-significant bits. In such situations, it is sufficient for the base-station to reconstruct the data of the node $i$ if the node $i$ sends only those bits where its data actually differs from that of node $j$.

Given the $B(X_i/X_j)$ number of its least-significant bits that the node $i$ has to send to the base-station, there are $2^{B(X_i/X_j)}$ possible bit-patterns, one out of which the node $i$ has to communicate to the base-station. Assuming that the each of these $2^{B(X_i/X_j)}$ bit-patterns are uniformly distributed with probability $1/2^{B(X_i/X_j)}$, then following the typical Huffman-coding argument, the node $i$ can communicate its data to the base-station in at most $B(X_i/X_j)$ bits on average.

Given a communication schedule $\pi$, the first node in the schedule $\pi(1)$ sends its $n$ bits to the base-station. Based on this and the knowledge of the correlation model given in \eqref{eqn:cor}, the base-station informs the second node $\pi(2)$ in $\lceil \log B(X_{\pi(2)}/X_{\pi(1)}) \rceil$ bits to send its information in $B(X_{\pi(2)}/X_{\pi(1)})$ bits. Assuming the uniform-distribution, the node $\pi(2)$ sends the requested information in $B(X_{\pi(2)}/X_{\pi(1)})$ bits, in average. Continuing this process, the base-station queries all the $N$ nodes. So, the average-case communication complexity $\bar{C}_{\pi}$ of schedule $\pi$ is the sum of the total number of bits sent by the base-station and the number of bits sent, on average, by the sensor-nodes. It is given by:
\begin{eqnarray}
\label{eqn:average_case}
\bar{C}_{\pi} & = & \lceil \log B(X_{\pi(2)}/X_{\pi(1)}) \rceil + \ldots \nonumber \\
          & & + \lceil \log B(X_{\pi(N)}/X_{\pi(1)}, \ldots, X_{\pi(N-1)}) \rceil \nonumber \\
          & & + B(X_{\pi(2)}/X_{\pi(1)}) + \ldots \nonumber \\
          & & + B(X_{\pi(N)}/X_{\pi(1)}, \ldots, X_{\pi(N-1)})
\end{eqnarray}

The optimal value of $\bar{C}_{\pi}$ is achieved for that schedule $\pi \in \Pi$ that solves the following optimization problem:
\begin{equation}
\label{optprob2}
\bar{C}_{min} = \min_{\pi \in \Pi} \bar{C}_{\pi}
\end{equation}

It is easy to see that, once again, the optimum schedule is generated by a greedy scheme that chooses the next node in the schedule to be that node that is nearest to the set of the nodes which are already scheduled.

\textit{Remark:} Comparing \eqref{eqn:worst_case} and \eqref{eqn:average_case}, it may appear that the average-case performance of a schedule is no better than its worst-case performance, but it should be noted that the average-case analysis is done for the uniform distribution of the bit-patterns, that gives the maximum entropy. For any other distribution of the bit-patterns, the average communication complexity will be lesser than $\bar{C}_{\pi}$ given by \eqref{eqn:average_case}.

\section{Conclusions and Future Work}
\label{conclusions}
This paper proposes a new framework, based on exploiting the redundancy induced by the spatio-temporal correlations in the sensor data and the reduction in communication complexity due to interaction, to reduce the total number of bits sent by the sensor nodes in a single-hop data-gathering sensor network. The proposed formalism views the problem of many sensor nodes communicating with the base-station as the problem of many informants with the correlated information communicating with single recipient. We extend various existing results on single informant-recipient pair communication to the present case. We show that such extensions lead to various non-trivial new results. Finally, we apply this new framework to compute the worst-case and average-case communication complexities of a typical sensor network scenario and demonstrate the significance of our contribution.
\newpage
Major contributions of our work are the following:
\begin{enumerate}
\item We show that interaction helps in reducing the communication complexity also in the scenarios where more than one informant are involved.
\item We show that when the \textit{ambiguity} is more than two, then no fixed number of messages are optimum.
\item We show that for the scenarios involving more than one informant, the order in which the informants communicate with the recipient may determine the communication complexity. We conjecture that if the nodes which have `more' information communicate first, then this brings down the overall communication complexity. Essentially, we need a metric to quantify the 'amount of information', but not in `Shannon-sense'.
\item We show that when multiple informants communicate with a recipient, the m-message complexity of communication between informant $Y$ and recipient $X$, can be computed by directly modifying the hypergraph $G(Y|X)$ based on the information provided by the previous informants in the communication schedule.
\end{enumerate}

In future, we would like to give more formal and detailed exposition of our work presented in this paper.

\end{document}